\documentclass[twocolumn]{aa} 
\usepackage{graphicx}
\usepackage{txfonts}
\begin{document}
  \title{Impact of the physical processes in the modeling of HD49933}

  \author{L. Piau
         \inst{1},
	  S. Turck-Chi\`eze \inst{1}
         V. Duez \inst{1},
	  \and R. F. Stein \inst{2}}

  \institute{CEA-Saclay, DSM/IRFU/SAp, L'Orme des merisiers, Batiment 709, 91191 Gif-sur-Yvette, France \\
             \email{laurent.piau@cea.fr}         \and
            Michigan State University,Department of Physics \& Astronomy
	     East Lansing, MI 48824-2320,
	     USA\\
            }



 \abstract
  {On its asteroseismic side, the initial run of CoRoT was partly devoted to the solar like star HD49933.
   The eigenmodes of this F dwarf have been observed with unprecedented accuracy.}
  {We investigate quantitatively the impact of changes in the modeling parameters 
  like mass and composition. More importantly we investigate how a sophisticated physics 
  affects the seismological picture of HD49933. 
  We consider the effects of diffusion, rotation and the changes in convection efficiency.}
  {We use the CESAM stellar evolution code coupled to the ADIPLS adiabatic pulsation package
   to build secular models and their associated oscillation frequencies. We also exploited the hydrodynamical 
   code STAGGER to perform surface convection calculations. The seismic variables used in this work are:
   the large frequency separation, the derivative of the surface phase shift, 
   and the eigenfrequencies $\rm \nu_{\ell=0,n=14}$ and $\rm \nu_{\ell=0,n=27}$. 
   }
  {Mass and uncertainties on the composition have much larger impacts on the seismic variables we consider
   than the rotation. The derivative of the surface phase shift is a promising variable for the determination
   of the helium content. The seismological variables of HD49933 are sensitive to the assumed
   solar composition and also to the presence of diffusion in the models.}
  {}


  \maketitle
%

\section{Introduction}\label{sec1}

The accurate measurements of oscillation frequencies by CoRoT (Baglin et al. 2006) in nearby solar-like stars such
as HD49933 (Appourchaux et al. 2008) aims at constraining their internal dynamics. 
First, the global parameters of the star (age, mass, composition) must be determined. For this,
the seismic data complement the information that already exists.
Second, a comparison with sophisticated models is required.
In this work we use an updated version of the CESAM stellar evolution code 
(Morel 1997). CESAM can take into account 
microscopic diffusion and the radiative accelerations of heavy elements. It 
also optionally includes the secular 
effects of rotation in the radiation zones (Mathis \& Zahn 2004). The resulting
mixing slightly changes the temperature profile of otherwise stably stratified regions and 
modifies the structure and evolution of the star (Maeder \& Meynet 2000, Decressin et al. 2009).   

Like helioseismology for the Sun, asteroseismology can constrain the helium 
fraction in solar analogs. In this respect stars slightly more
massive than the Sun are especially interesting targets. Asteroseismology 
can also constrain convection: the depth of the outer 
convection zone depends on the efficiency of convective energy
transport in the subsurface layers and in turn affects the seismic signal.
In the current work, we compute the relation between the effective temperature and
the specific entropy in the deep convection zone where the motions become adiabatic. 
To this purpose, we use the STAGGER hydrodynamical code (Stein \& Nordlund 1998).
This computation gives an idea of the magnitude of the 
change of the mixing length main parameter $\rm \alpha_{mlt}$ with the effective temperature 
for stars having surface conditions close to the solar ones like HD49933.

Although some of our models are in fair agreement with 
the current classical and seismological observations of HD49933, 
our purpose is not to find the model that best fits them. We intend to 
estimate the influence of mass, composition and of different
internal processes on the seismic indicators and the surface
properties of HD49933. Section \ref{sec2} presents our secular models 
of HD49933 and explores the impact of mass, composition, microscopic diffusion and 
rotation. Section \ref{sec3} presents the effects of a modified surface convection efficiency
and conclusion follows in section \ref{sec4}.


\section{Secular models of HD49933}\label{sec2}

We build models of HD49933 using the CESAM stellar evolution code (Morel 1997). The models are
initiated on the zero age main sequence. We use the NACRE compilation of nuclear reaction rates  
(Angulo et al. 1999). The equation of state (OPAL2001) and high temperature 
opacities come from the OPAL group (Rogers, Swenson \& Iglesias 1996, Iglesias \& Rogers 
1996). Below 5800 K we use the opacities of Ferguson et al. (2005).
The equation of state and opacity tables were computed for the recent solar metal repartition
advocated by Asplund et al. (2005). The atmosphere is buildt
using the Hopf law (Mihalas 1978). It is connected to the envelope
at the optical depth 20.
Finally, the convection zone is fully homogeneous and modeled using mixing-length theory
(hereafter MLT) in a formalism close to B\"ohm-Vitense (1958) (see Piau et al. 2005 for the exact
description of the current MLT formalism).
Our Sun calibrated value for the main MLT parameter is $\alpha_\mathrm{MLT}=1.587$.

Solano et al. (2005) find [Fe/H]=-0.37. If we assume the Asplund et al. (2005) solar surface composition 
(X=0.7392, Z=0.0122) and metal repartition then HD49933 should currently exhibit X=0.7664, $\rm Z=5.4\,10^{-3}$
provided the helium and metals surface abundances follow the typical
Galactic enrichment law $\rm \frac{\Delta Y}{\Delta Z}=3$ (Fernandes et al. 1998)
\footnote{ It should be noted that the absolute Z/X=$7.04\,10^{-3}$ we adopt 
is smaller than the one adopted in the study of Goupil et al. (2009) Z/X=$0.0106$.
Unlike us these authors consider the solar metal abundance and 
mixture of Grevesse \& Noels (1993).}.
The effective temperature found in the literature ranges from $\rm T_{eff}=6780 \pm 130 K$
(Bruntt et al. 2008) to $\rm T_{eff}=6700 \pm 65 K$ (Gillon \& Magain 2006)
but values as low as $\rm T_{eff}=6600 \pm 130 K$ have also been suggested
(Lastennet et al. 2001). For a visual magnitude of $\rm m_v=5.77$, the HIPPARCOS parallax 
($\rm 33.7 \pm 0.4 mas$, van Leeuwen 2007) and bolometric correction ($\rm BC=0.025 \pm 0.005$, 
Bessell et al. 1998) suggest $\rm log (L/L_{\odot})=0.53 \pm 0.01$. The following models
have been calibrated in luminosity in the sense that we stop 
the evolution once they reach the observed luminosity.
Unless explicitly mentioned 
all the models include 
overshooting of $0.2$ pressure scale height (hereafter $\rm H_p$) that is typical 
of the expected mass range of the star (Claret 2007).
We investigate the impact of the mass, the composition and 
the rotation on the seismic properties of the star.
The oscillation frequencies are computed using the Aarhus adiabatic pulsation
package ADIPLS (Christensen-Dalsgaard \& Berthomieu 1991).
We focus our attention on three seismic quantities: the large 
frequency separation ($\Delta \nu = \nu_{\ell,n+1}-\nu_{\ell,n}$) 
that is sensitive to the global properties
of the star such as mass and age, the lowest degree lowest and highest order modes
frequency currently observed ($\rm \nu_{\ell=0,n=14}$ and $\rm \nu_{\ell=0,n=27}$)
and the derivative of the surface 
phase shift with frequency ($\beta^{\star} (\nu)$) in order
to address the helium ionization effects. We do not investigate the small 
separation or other seismic variables.
To determine the averages of these quantities, we excludes frequencies below 
a cut off of 1mHz in order to be in the asymptotic regime.

\subsection{Mass effects}\label{sec21}

HD49933 being a single star, its mass estimate relies on evolutionary tracks. 
It is therefore sensitive not only to the approach chosen in the modelling
(e. g. description of diffusion, rotation mixing, overshooting... ) but 
also to the input parameters of the models such as the metallicity and
the helium fraction.
For these reasons it is interesting to address the 
influence of the mass on the seismic signature of this star.
Mosser et al. (2005) suggested $\rm M/M_{\odot} \approx 1.2$ and we 
explore the effect of mass variation around this value. 
No diffusion or rotation effects are taken into account
in the models of this subsection. The convection
overshooting in the core is $\rm 0.2 H_p$, however the 
overshooting has a very moderate impact on the variables
we consider here. As illustrated by the work of Goupil et al.
(2009), overshooting mostly influences the small seismic 
separation which we have not considered here. As an indication
of the overshooting impact, we provide
a 1.17 $\rm M_{\odot}$ model without overshooting on the 
last line of Table \ref{tab1}.

\begin{figure}[Ht]
\includegraphics[width=8.cm,clip]{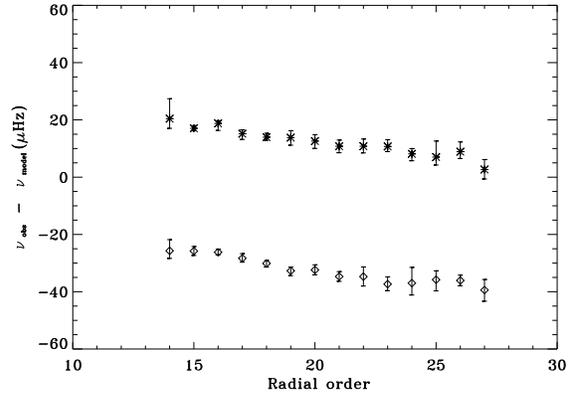}
\caption{Differences between the observed eigenfrequencies and the model eigenfrequencies
vs. the radial order for degree modes $\ell=$0. 
The model considered is model  3, losanges and stars respectively 
correspond to the $\rm M_A$ and $\rm M_B$ modes identifications of Benomar et al. (2009). 
We also indicated the $\rm 2 \sigma$ error bars from these authors.}
\label{11947fg1}
\end{figure}

Table \ref{tab1} shows the values of the large separation and the 
frequencies of the modes of degree $\ell$=0 and radial orders n=14 and n=27.
These modes should respectively minimize and maximize the surface effects.
The recent analysis of Benomar et al. (2009) suggests two slightly different
observational solutions owing to the modes identification. The first
mode identification, named $\rm M_A$ by Benomar et al.,  
corresponds to the work of Appourchaux et al. (2008) $\rm \Delta \nu = 85.92 \pm 0.43 \mu Hz$, 
$\rm \nu_{\ell=0,n=14}=1244.43 \pm 3.90 \mu Hz$ and $\rm \nu_{\ell=0,n=27}=2363.81 \pm 3.90 \mu Hz$.
The other mode identification, $\rm M_B$, provides $\rm \Delta \nu = 85.81 \pm 0.29 \mu Hz$, 
$\rm \nu_{\ell=0,n=14}=1290.55 \pm 6.97 \mu Hz$ and $\rm \nu_{\ell=0,n=27}=2405.90 \pm 3.48 \mu Hz$
($2 \sigma$ error bars).
The models in Table \ref{tab1} all have the same composition: 
X=0.7664, $\rm Z=5.4\,10^{-3}$. A 0.2 $\rm H_p$ core convection 
overshooting is included in all models except the one mentioned 
by $\dagger$ where there is no overshooting. The second part of 
the table displays the current seismic observations following the 
two possible mode identification $\rm M_A$ and $\rm M_B$ of Benomar et al. (2009).
The best fit with seismic data is obtained for $\rm 1.17 M_{\odot}$. Within error bars, the large separation of
this model agrees with identification $\rm M_A$ of Benomar et al. (2009)
and is marginaly above their identification $\rm M_B$. The absolute frequencies 
of our model and of this second solution agree at low and high order.
Figure \ref{11947fg1} displays the difference in eigenfrequencies between our
model and the two observational solutions for degree $\ell$=0,
as a function of the radial 
order.  The difference is always below $\rm 40 \mu Hz$.
Note that in the solar case, predicted and observed eigenmodes
frequencies agree at low frequencies but differ by $\rm \sim 20 \mu Hz$ 
at high frequencies most likely because of surface effects (Turck-Chi\`eze 
et al. 1997). The average difference between the model and the first and
second observed solutions of Benomar et al. (2009) are respectively $\rm -32.6 \mu Hz$ and
$\rm 12.2 \mu Hz$. In both cases there is a clear trend in the model/observations difference
with increasing frequency which might be due to the absence of subphotospheric 
turbulence and/or magnetic surface effects as in the solar case.

\begin{table}
  \caption[]{Mass effect on the large frequency separation, $\rm \nu_{\ell=0,n=14}$ and $\rm \nu_{\ell=0,n=27}$.}
     \label{tab1}
 $$ 
     \begin{array}{lllllc}
        \hline
        \noalign{\smallskip}
        $$\rm M/M_{\odot}$$       &  $$\rm T_{eff}$$ & \Delta \nu           &   \nu_{\ell=0,n=14}     &   \nu_{\ell=0,n=27}    & {\rm Model} \\
                                  & {[\mathrm{K}]}   & {[\mathrm{\mu Hz}]}  &   {[\mathrm{\mu Hz}]}   &   {[\mathrm{\mu Hz}]}  &               \\
        \noalign{\smallskip}
        \hline
        \noalign{\smallskip}
        1.25                   & 6924                    &   100.1                                      &  1480                                 &        2780              &  1  \\
        1.2                    & 6746                    &    92.1                                      &  1359                                 &        2561              &  2  \\
        1.17                   & 6610                    &    86.2                                      &  1270                                 &        2403              &  3  \\
        1.17  $$\dagger$$      & 6638                    &    87.1                                      &  1280                                 &        2424              &  3bis  \\
        \noalign{\smallskip}
        \hline

        \noalign{\smallskip}
        $$\rm Mode \,identification$$   &                &                                              &                                       &                          &  \\
        \noalign{\smallskip}
        \hline
        \noalign{\smallskip}
        $$\rm M_A$$                &                    &   85.9                                      &  1244                                 &        2364              &   \\
        $$\rm M_B$$                &                    &   85.8                                      &  1290                                 &        2406              &   \\
        \noalign{\smallskip}
        \hline

     \end{array}
 $$ 
\end{table}

The large separation, $\rm \nu_{\ell=0,n=14}$ and $\rm \nu_{\ell=0,n=27}$ 
decreases rapidly with mass and with current modeling assumptions 
suggests a mass slightly below 1.2 $\rm M_{\odot}$.
The logarithmic derivative of the large separation
with mass around $\rm 1.17 M_{\odot}$ is $\rm \frac{\partial ln \Delta \nu}{\partial ln \Delta M}=2.6$
which means a significant dependence of $\Delta \nu$ on mass (see hereafter).
If we extrapolate this result to the average large separation observed 
by Benomar et al. (2009) $\rm 85.92 \mu Hz$, we find that
the mass of HD49933 should be $\rm  1.168 M_{\odot}$. We note 
that the effective temperature of our $\rm 1.17 M_{\odot}$ 
model is between 100 and 150 K below the current estimates
but in agreement with the previous determination of Lastennet et al.(2001).
We recall that this model was obtained assuming [Fe/H]=-0.37,
the solar Asplund et al. (2005) metal repartition and $\rm \frac{\Delta Y}{\Delta Z}=3$ 
as the metal vs. helium relation. In the next section
we study the effects of composition.

\subsection{Composition effects}\label{sec22}

As recent developments in solar physics have shown, helioseismological results
are very sensitive to the solar composition and therefore stand
as key indicators of it (Turck-Chi\`eze et al. 2004, 
Bahcall et al. 2005). On the one hand the metals are the main contributors to
the opacities in low mass stars (Turck-Chi\`eze et al. 2009, Turck-Chi\`eze et al. 1993) which directly 
affects the structure and the seismic signal. On the other
hand, because of the low effective temperature, the helium fraction is only accessible through
seismology (Basu \& Antia 1995). Yet the mass and the age of stars other than the Sun have
to be known if one wants to constrain their metals or 
helium fractions (Basu et al. 2004). In the case of HD49933 we can therefore only give qualitative
indications on the composition.

We addressed the composition issue using the $\rm 1.17 M_{\odot}$ models, as the mass 
effects suggest this is the best fit for HD49933.
The A05 and GS98 respectively stand for Asplund et al. (2005) and 
Grevesse \& Sauval (1998) solar metal repartitions. In both cases the assumed metallicity is [Fe/H]=-0.37 dex
but in the former case this corresponds to a smaller  absolute amount of metals
other than iron (mostly oxygen and carbon) than in the later case. If we assume the A05 metal repartition
then X=0.7664 and $\rm Z=5.4\,10^{-3}$ (model  3) whereas if we assume the GS98 
metal repartition X=0.7638, $\rm Z=8.02\,10^{-3}$ (model  4). These are the current {\it and} initial 
convection zone compositions because no microscopic diffusion is taken into account in 
this subsection.
As illustrated in Table \ref{tab2} the metal repartition and helium abundance  
have significant impacts on $\Delta \nu$ although smaller than the impact of mass (\S \ref{sec21}).
Please note that the models in Table \ref{tab2} all have the same mass: $\rm 1.17 M_{\odot}$.
The main difference between A05 and GS98 metal repartition is the 
[O/Fe] ratio. The logarithmic derivative of the large separation
with [O/Fe] is $\rm \frac{\partial ln \Delta \nu}{\partial ln \Delta [O/Fe]}=0.83$
\footnote{Asplund et al. (2005) provide [O/Fe]=1.21 for the Sun and a variation of [O/Fe]
of $\approx -0.2$ with respect to Grevesse \& Sauval (1998) earlier work.}.
We mention that variations of the [O/Fe] ratio are observed in nearby stars,
the scatter of [O/Fe] across the Galactic disk being the
order of 0.2 dex (Edvardsson et al. 1993). 

The decrease in the global metal content
from GS98 to A05 creates a drop in opacity and therefore a somewhat higher
effective temperature at a given luminosity. This corresponds 
to a smaller radius and a higher interior temperature and is
consistent with a larger large separation as:
$$\frac{1}{2 \Delta\nu} \sim \int_{0}^{R} \frac{dr}{c}$$
Thus more metal rich models (GS98) suggest a mass higher than 
$\rm 1.17 M_{\odot}$. Quantitatively, going from the 
A05 to the GS98 composition lowers the large difference by
$\rm \sim 8 \mu$ Hz which is comparable to the mass effect between
models  2 and  3 ($\rm 0.03 M_{\odot}$ difference).


\begin{table}[Ht]
  \caption[]{Composition effects on the large frequency separation, $\rm \nu_{\ell=0,n=14}$ and $\rm \nu_{\ell=0,n=27}$.}
     \label{tab2}
 $$ 
     \begin{array}{p{0.15\linewidth}llllc}
        \hline
        \noalign{\smallskip}
        Metal        &  $$\rm T_{eff}$$ & \Delta \nu           &   \nu_{\ell=0,n=14}     &   \nu_{\ell=0,n=27}    & {\rm Model\,} \\
        repartition  & {[\mathrm{K}]}   & {[\mathrm{\mu Hz}]}  &   {[\mathrm{\mu Hz}]}   &   {[\mathrm{\mu Hz}]}  &                 \\
        \& $\frac{\Delta Y}{\Delta Z}$           \\
        \noalign{\smallskip}
        \hline
        \noalign{\smallskip}
        A05      \hspace{1.8cm}    $\frac{\Delta Y}{\Delta Z}=3$                 & 6610                    &    86.2            &  1270                                &   2403         & 3 \\
	 \hline
        \noalign{\smallskip}
        GS98     \hspace{1.6cm}    $\frac{\Delta Y}{\Delta Z}=3$                 & 6230                    &    74.4            &  1160                                &   2139         & 4 \\
	 \hline
        \noalign{\smallskip}
        A05      \hspace{1.8cm}    $\frac{\Delta Y}{\Delta Z}=1$                 & 6742                    &    90.8            &  1338                                &   2525         & 5 \\
	 \hline
        \noalign{\smallskip}
     \end{array}
 $$ 
\end{table}

\begin{figure}[Ht]
\includegraphics[width=8.cm,clip]{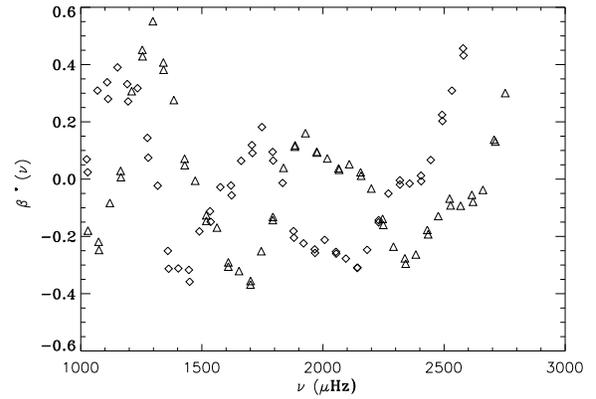}
\caption{Losanges: $\beta^{\star} (\nu)$ vs. $\nu$ for the model having 
$\rm M_{\star}=1.17 M_{\odot}$ and X=0.7664, Y=0.2282 and $\rm Z=5.4\,10^{-3}$ (model  3).
Triangles: $\beta^{\star} (\nu)$ vs. $\nu$ for the same mass but X=0.7529, Y=0.2417 and $\rm Z=5.3\,10^{-3}$ (model  5).}
\label{11947fg2}
\end{figure}

\begin{figure}[Ht]
\includegraphics[width=8.4cm,clip]{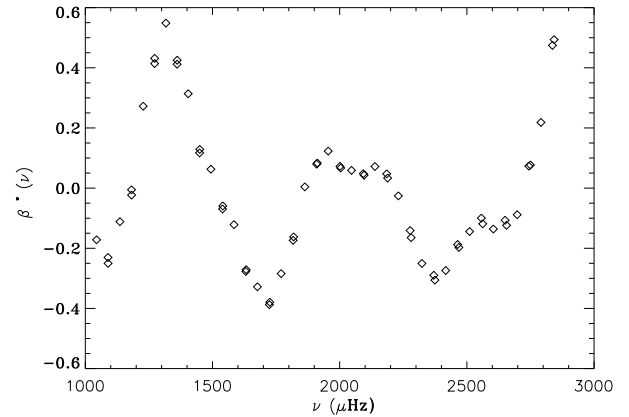}
\caption{$\beta^{\star} (\nu)$ vs. $\nu$ for the model having $\rm M_{\star}=1.2 M_{\odot}$,  X=0.7664, Y=0.2282 and $\rm Z=5.4\,10^{-3}$ (model  2).}
\label{11947fg3}
\end{figure}

The region of the helium second ionization, located right below 
the photosphere, appears as a discontinuity to the
internal waves. As such it creates an oscillatory pattern in the eigenmodes
frequencies (Gough 1990) that has been used for some time 
to infer, for instance, the solar photosphere abundance in helium (Vorontsov et al. 1991).
The oscillatory pattern is clearly visible in the 
second difference seismic variable $\delta_2 \nu_{\ell,n}= \nu_{\ell,n-1} -2\nu_{\ell,n} + \nu_{\ell,n+1}$
especially in stars slightly more massive than the Sun (Piau et al. 2005).
While $\delta_2 \nu_{\ell,n}$ is probably easier to determine
from the observations than the surface phase shift $\alpha (\nu)$, this later variable
and its derivatives are less sensitive to uncertainties in frequency determinations (Lopes et al. 1997).

Following Lopes et al. (1997) we compute $\beta (\nu)= \alpha (\nu) - \nu \frac{d \alpha}{d \nu}$ to extract the effect of helium
in the cases of three stars: models 2, 3 and 5 (see Tables \ref{tab1} and \ref{tab2}). 
$\beta (\nu)$ is estimated for the low degree modes $\ell=0,1,2$.
Then it is corrected by a linear fit in order to obtain $\beta^{\star} (\nu)$
whose mean value is 0. $\beta^{\star} (\nu)$ vs. $\nu$ is shown on figures \ref{11947fg2} and
\ref{11947fg3}. The differences between models 2 and
3 account for the effects of mass whereas the differences between the models 3 and
5 account for the effect of the helium fraction. Models 2 and 3 have
an helium mass fraction of Y=0.2282. Because of a different $\frac{\Delta Y}{\Delta Z}$ law
model 5 is richer in helium than the other models with Y=0.2417. 
Note that all our models have the same metallicity [Fe/H]=-0.37.

Figure \ref{11947fg2} illustrates the strong dependence of the peaks of $\beta^{\star} (\nu)$ and
therefore $\beta (\nu)$ on the helium mass fraction. The increase of
$\rm \Delta Y$=0.0135 from model  3 to  5 increases the frequency of the first peak of 
$\beta^{\star} (\nu)$ above 1 mHz by $\approx$  $\rm 150 \mu Hz$. The second peak of 
$\beta^{\star} (\nu)$ is shifted 
by $\approx$ $\rm 200 \mu Hz$. However Figure \ref{11947fg3} shows that
the position of the peaks in $\beta^{\star} (\nu)$ is also strongly related to the mass.
Based on the differences from model  3 to  5, the logarithmic derivative of the large separation
with the helium mass fraction is $\rm \frac{\partial ln \Delta \nu}{\partial ln \Delta Y}=0.90$.

\subsection{Diffusion effects}\label{sec23}

The consequences of microscopic diffusion in the enveloppes of solar-like stars
are clearly observed. In the case of the Sun it has been known for 
some time that it improves the agreement between theoretical and observed 
sound speed profiles (Christensen-Dalsgaard 
et al. 1993) and it is a necessary process to explain the current photospheric helium
abundance (Basu \& Antia 1995). Besides this, diffusion is the best candidate to 
explain the $^7$Li abundance in Population II solar analogs (Piau 2008, Richard et al. 2005).
Finally refined calculations by Turcotte et al. (1998) have shown that
a significant change in surface composition through diffusion is expected 
between $\rm 1.1 M_{\odot}$ and $\rm 1.5 M_{\odot}$ (the typical mass range of HD49933). We calculate 
the joint effects of microscopic diffusion (following the formalism of
Burgers 1969) and radiative accelerations.  We consider that the initial
composition is the initial composition of our calibrated solar
models including diffusion i.e. X=0.7195 and Y=0.2664 (Turck-Chi\`eze et al. 2004).

Because diffusion takes the metals away from the surface, the
diffusive models for a given mass and $\rm Z/X$ surface ratio are
cooler than the non diffusive models with the same $\rm Z/X$ surface ratio.
For instance a $\rm 1.25 M_{\odot}$ non diffusive model with $\rm Z/X=1.15\,10^{-2}$ and $\rm Y=0.2664$ 
has an effective temperature
of $\rm T_{eff}=6856 K$ when it reaches the luminosity $\rm log(L/L_{\odot})=0.53$ at 1.12 Gyr. 
The corresponding $\rm 1.25 M_{\odot}$
diffusive model exhibits $\rm T_{eff}=6151 K$ at the same luminosity
which it reaches at 3.32 Gyr (see Table \ref{tab3}).
Because of this effect, the diffusive models complying with
the observed effective temperature of HD49933 are more massive than the 
non diffusive ones. Table \ref{tab3} shows that they
lie around $\rm 1.35 M_{\odot}$ while the non diffusive model with 
the correct effective temperature have masses around $\rm 1.2 M_{\odot}$.
Regarding the seismic properties of diffusive models the same
remark can be done. The diffusive model that best agrees with the observations
is significantly more massive ($\rm 1.3 M_{\odot}$) than the non diffusive one 
($\rm 1.17 M_{\odot}$). Being more massive than the non diffusive
models, the diffusive models fitting the observations are also 
younger: the $\rm 1.3 M_{\odot}$ diffusive model is 1.79 Gyr 
whereas the $\rm 1.17 M_{\odot}$ non diffusive model is 3.59 Gyr.

As stated above, if we assume the Asplund et al. (2005) solar 
composition, then in HD49933 we have $\rm Z/X=7.04\,10^{-3}$.  
Table \ref{tab3} shows that the diffusive models that meet this
condition are too cool, or equivalently that the diffusive models
in the right range of the effective temperature are too metal
poor. The physical reason for this being that in
the higher mass regime the outer convection zone 
becomes too shallow for convection to brake significantly 
the gravitational settling of the heavy elements. This drawback suggests that the microscopic diffusion 
(and radiative acceleration) effects are moderated by
an additional mixing process in the radiation envelope 
below the outer convection zone. There is actually a strong
evidence of such a mixing as, with $\rm T_{eff}=6780 K$ HD49933, nearly stands to the 
middle of the so-called lithium dip (Boesgaard \& Tripicco 1986, Boesgaard \& King 2002).
The stars of the lithium dip show lithium and beryllium depletion 
which implies deep mixing in their radiation zones. A likely 
candidate to this is the shear turbulence induced by 
the angular momentum loss (Talon \& Charbonnel 1998, Decressin et al. 2009). 
The occurance of such a process within HD49933 would brake the microscopic 
diffusion. Similarly the tachocline mixing (Spiegel \& Zahn 1992) in the 
transition region between convection and radiation zones brakes the 
metals and helium diffusion in the Sun (Turck-Chi\`eze et al. 2004). The interplay 
between  diffusion and this process -or another non standard mixing phenomenon-
appears necessary to build the observed $^7$Li abundance in Population II 
(Piau 2008, Richard et al. 2005). In the next subsection 
we briefly address the effects of rotation and of 
the interaction between diffusion and rotation.

\begin{table}[Ht]
  \caption[]{Microscopic diffusion and radiative acceleration effects on the large frequency 
separation, $\rm \nu_{\ell=0,n=14}$ \& $\rm \nu_{\ell=0,n=27}$.}
     \label{tab3}
 $$ 
     \begin{array}{p{0.17\linewidth}llllllc}
        \hline
        \noalign{\smallskip}
        $\rm M/M_{\odot} $        &  $$\rm T_{eff}$$ &  $$\rm Z/X$$  & \Delta \nu           &   \nu_{\ell=0,n=14}     &   \nu_{\ell=0,n=27}    & {\rm Model\,} \\
                                  & {[\mathrm{K}]}   &               & {[\mathrm{\mu Hz}]}  &   {[\mathrm{\mu Hz}]}   &   {[\mathrm{\mu Hz}]}  &                 \\
        \noalign{\smallskip}
        \hline
        \noalign{\smallskip}
        1.35                      & 6668                    &   $$4.1\,10^{-3}$$   &  94.0                            & 1401  & 2627             & 6 \\
        1.3                       & 6480                    &   $$4.2\,10^{-3}$$   &  85.7                            & 1280  & 2401             & 7 \\
        1.25                      & 6151                    &   $$1.15\,10^{-2}$$  &  76.3                            & 1216  & 2209             & 8 \\
        \noalign{\smallskip}
        \hline
     \end{array}
 $$
\note{The models all have the same initial composition (X=0.7195, Y=0.2664
and thus $\rm Z/X=1.959\,10^{-2}$) and are stopped at the same luminosity ($\rm log L/L_{\odot}=0.53$).
Then the ages of the 1.25, 1.3 and 1.35 $\rm M_{\odot}$ models are respectively 3.32, 1.79 and 0.54 Gyr.}
\end{table}

\subsection{Rotation effects}\label{sec24}

The power spectrum of HD49933 suggests a surface rotation period of 
3.4 days (Appourchaux et al. 2008). If, following Th\'evenin
et al. (2006), we consider $\rm R_{\star}/ R_{\odot}=1.43$
this leads to an equatorial velocity $\rm V_{eq}= 21.3\,km s^{-1}$.
This result is in agreement with direct estimates of $\rm v sin i$  
(Solano et al. 2005, Mosser et al. 2005). 

We have applied the theory of Mathis \& Zahn (2004) for 
two $\rm 1.17 M_{\odot}$ models evolving without angular momentum loss and showing
equatorial velocities $\rm V_{eq}=10\,km s^{-1}$ (model  9) and $\rm V_{eq}=20\,km s^{-1}$ 
(model 10) at 4.35 Gyr and 3.97 Gyr respectively. 
These ages correspond to $\rm log (L/L_{\odot}) = 0.53$.
The Mathis \& Zahn (2004) transport equations account for
the modifications in temperature, mean molecular 
weight and gravitational potential induced by rotation.
The meridional circulation is calculated as well.
Table \ref{tab4} provides information similar to the preceeding ones.
The model 11 mentionned with $\dagger$ includes both rotation
and microscopic diffusion. Its initial composition is
X=0.7195, Y=0.2664 ($\rm Z/X= 1.959 \, 10^{-2}$). All the
other models of the Table have rotation but no microscopic
diffusion effects and an initial composition X=0.7664, $\rm Z=5.4\,10^{-3}$.
The impact of rotation effects on the large frequency separation is hardly visible.
The logarithmic derivative of the large separation
with the equatorial velocity computed from models 9 and 10
is tiny $\rm \frac{\partial ln \Delta \nu}{\partial ln V_{eq}}=1.1\, 10^{-3}$.

Thus the rotation effects seem negligible. However when tested in diffusive models
they diminish the diffusion efficiency and therefore help
improving the issue outlined in \S \ref{sec23}.
The physics used in building models  7 and  11 is similar except for rotation.
Table \ref{tab4} shows that they nearly are identical except their
difference in the surface $\rm Z/X$ ratio. Model  11 is
not our best fit to the observational constrains but it is
the most sophisticated secular model of this work as it 
includes both of the significant diffusion and rotation effects
expected in HD49933. Its large frequency separation 
and surface metallicity lie very close to the observations.
Even tough model  11 is too cool, it points out
that the mass of HD49933 predicted by refined models is higher 
than the mass predicted by simple models (e.g. model 3).

\begin{table}[Ht]
  \caption[]{Rotation and microscopic diffusion effects on the large frequency 
separation, $\rm \nu_{\ell=0,n=14}$ \& $\rm \nu_{\ell=0,n=27}$.}
     \label{tab4}
 $$ 
     \begin{array}{lllllllc}
        \hline
        \noalign{\smallskip}
        $$\rm M/M_{\odot} $$   &  Z/X      & $$\rm T_{eff}$$  & \Delta \nu           &   \nu_{\ell=0,n=14}   &  \nu_{\ell=0,n=27}    & $$\rm V_{eq}$$         & {\rm Model\,} \\
                               &           & {[\mathrm{K}]}   & {[\mathrm{\mu Hz}]}  &   {[\mathrm{\mu Hz}]} &  {[\mathrm{\mu Hz}]}  & $$[\rm{km.s^{-1}}]$$   &                \\
        \noalign{\smallskip}
        \hline
        \noalign{\smallskip}
        1.17                   &  $$7.04\,10^{-3}$$  & 6610                    &    86.2                        &  1270        & 2403       & 0.0       & 3 \\
        1.17                   &  $$7.04\,10^{-3}$$  & 6614                    &    86.2                        &  1283        & 2410       & 10.       & 9 \\
        1.17                   &  $$7.04\,10^{-3}$$  & 6613                    &    86.1                        &  1282        & 2408       & 20.       & 10\\
        1.30                   &  $$7.04\,10^{-3}$$  & 6480                    &    85.7                        &  1280        & 2401       & 0.0       & 7 \\
        1.30   $$\dagger$$     &  $$5.28\,10^{-3}$$  & 6480                    &    85.7                        &  1281        & 2401       & 20.       & 11\\
        \noalign{\smallskip}
        \hline
     \end{array}
 $$ 
\end{table}
\section{Surface convection of HD49933}\label{sec3}

\subsection{Dynamical models}\label{sec31}

We use the STAGGER code (Stein \& Nordlund 1998) to investigate 
the convective and the radiative energy transfer 
from 500 km above the photosphere downto 2500 km below. The computational 
domain extends over 6000 by 6000 km horizontaly. The current grid has 63 points in each direction. 
We solve the fully compressible equations of hydrodynamics
and use the same equation of state -OPAL- as in the CESAM calculations.
As far as possible, we use the same opacities as well.
Near and above the photosphere, the diffusion approximation for radiative energy
transport breaks down and one has to solve the radiative energy transfer 
equation. We adopted the binning method to solve this equation (Nordlund 1982).
This widely used method (see e.g. Ludwig et al. 2006) requires 1D atmosphere model structure 
as well as monochromatic opacities. For both we use the Kurucz data and programs (Castelli 2005a, 2005b).
Thus the opacities near and above the photosphere are not the OPAL opacities.
The equation of state and opacity tables where computed for
the composition  X=0.7664, $\rm Z=5.4\,10^{-3}$ (see \S \ref{sec22}).

We run STAGGER over the typical Kelvin-Helmholtz time-scale of the 
computational box that is $10^4$ s. The effective temperature is computed
using the surface radiative flux.
The efficiency of the surface convection is a priori dependent on the
total energy flux (i.e. the $\rm T_{eff}$) and the surface gravity.
We considered three simulations: a reference
one for the Sun (model A), a model where the specific entropy of the deep convection
zone was slightly increased with respect to the Sun (model B)
and a model where the surface gravity was slightly lowered (model C).
Model B is obtained by increasing the internal energy entering the
lower boundary of the domain by 2 \%. The gravity field in model B is kept
to its solar surface value and constant throughout the simulation box. Model C
is obtained by decreasing the surface gravity by
2 \% but the specific entropy of the deep convection zone
keeps its solar value of model A.

Table \ref{tab5} gives the surface conditions and the specific entropy in the deep convection 
zone (adiabatic regime) in the hydrodynamical models A, B, C and our stellar evolution code 
best fit model. The increase of the specific entropy in the adiabatic regime ($\rm s_{ad}$)
or the decrease of the surface gravity induce a decrease in the 
effective temperature.
Around the solar surface conditions, the logarithmic derivative of 
$\rm s_{ad}$ with $\rm T_{eff}$ is $\rm \frac{\partial ln s_{ad}}{\partial ln \Delta T_{eff}}=-0.225$
between models A and B.

If we extrapolate this to the case of HD49933 ($\rm T_{eff}=6780\,K$) it provides
$\rm s_{ad\,HD49933}= 1.8605\, 10^9erg.K^{-1}.g^{-1}$. Our best fit
model of HD49933 (model $ 3$ see Table \ref{tab1}) based on a 
solar calibrated value of $\rm \alpha_{mlt}$ exhibits $\rm s_{ad}= 2.3370\,10^9 erg.K^{-1}.g^{-1}$.
This suggests that the efficiency of convection increases
when going from the Sun to the HD49933 surface effective temperature (decreasing 
$\rm s_{ad}$ corresponds to increasing  $\rm \alpha_{mlt}$).
However the effects of changing the surface gravity are not accounted for.

\begin{table}[Ht]
  \caption[]{Surface gravity, effective temperature and deep convection zone specific entropy 
for the hydrodynamical surface convection models of HD49933 and the best fit model, model 3, 
computed with CESAM.}
     \label{tab5}
 $$ 
     \begin{array}{p{0.2\linewidth}cccc}
        \hline
        \noalign{\smallskip}
                                                            & \rm STAGGER        & \rm STAGGER           & \rm STAGGER         & \rm CESAM               \\
                                                            & \rm model\,A       & \rm model\,B          & \rm model\,C        & \rm model\,  3        \\
        \noalign{\smallskip}
        \hline
        \noalign{\smallskip}
        $\rm g [cm.s^{-2}]$                                 &  2.75\,10^4     &    2.75\,10^4     &    2.70\,10^4   &  1.618\,10^4         \\
        $\rm T_{eff} [K]$                                   &  5774           &    5639           &    5640         &  6610               \\
        $\rm s_{ad}  [erg.K^{-1}.g^{-1}]$                   &  1.9365\,10^9   &    1.9462\,10^9   &    1.9363\,10^9 &  2.3370\,10^9       \\
        \noalign{\smallskip}
        \hline
     \end{array}
 $$ 
\end{table}

\subsection{The mixing length parameter}\label{sec32}

Our hydrodynamical models suggest that
the efficiency of convection is higher in HD49933 than in
the Sun. This means that the $\rm \alpha_{mlt}$ adopted for
HD49933 should be larger. Table \ref{tab6} estimates the
effects of a modified $\rm \alpha_{mlt}$ on $\rm \Delta \nu$, $\rm \nu_{\ell=0,n=14}$
and $\rm \nu_{\ell=0,n=27}$ for $\rm 1.17 M_{\odot}$ models
with X=0.7664 and $\rm Z=5.4\,10^{-3}$.
As for the other models, core convection overshooting over $\rm 0.2 H_p$ is included.
The specific entropies $\rm s_{ad}$ of the deep convection zones
in models 8, 3 and 9 are respectively: 
$\rm 2.2962\,10^9 erg.K^{-1}.g^{-1}$, $\rm 2.3370\,10^9 erg.K^{-1}.g^{-1}$ and 
$\rm 2.4371\,10^9 erg.K^{-1}.g^{-1}$.
Our hydrodynamical models (\S \ref{sec31}) qualitatively
predict a typical change ($\rm s_{ad\,HD49933}-s_{ad\,Model A}$) 
in $\rm s_{ad}$ larger than what changes of 0.2 in $\rm \alpha_{mlt}$ would induce.
We conclude that a variation of 0.2 in $\rm \alpha_{mlt}$
between the Sun and HD49933 is plausible.

\begin{table}[Ht]
  \caption[]{Convection efficiency effect on the large frequency separation, $\rm \nu_{\ell=0,n=14}$ \& $\rm \nu_{\ell=0,n=27}$.}
     \label{tab6}
 $$ 
     \begin{array}{p{0.1\linewidth}llllc}
        \hline
        \noalign{\smallskip}
        $\rm \alpha_{mlt}$     &  $$\rm T_{eff}$$ {[\mathrm{K}]} & \Delta \nu {[\mathrm{\mu Hz}]} & \nu_{\ell=0,n=14} {[\mathrm{\mu Hz}]} & \nu_{\ell=0,n=27} {[\mathrm{\mu Hz}]}&  {\rm Model\,} \\
        \noalign{\smallskip}
        \hline
        \noalign{\smallskip}
        1.787                  & 6672                    &    88.9                        &  1316                                 &  2484                  &   8 \\
        1.587                  & 6610                    &    86.2                        &  1270                                 &  2403                  &   3 \\
        1.387                  & 6541                    &    83.5                        &  1471                                 &  2317                  &   9 \\
        \noalign{\smallskip}
        \hline
     \end{array}
 $$ 
\end{table}

The trend in $\rm \Delta \nu$ and frequency for an increasing efficiency
of convection disagrees with the observations. Benomar et al. (2009) favored
identification is $\rm \Delta \nu = 85.92 \pm 0.43 \mu Hz$, 
$\rm \nu_{\ell=0,n=14}=1244.43 \pm 3.90 \mu Hz$ and $\rm \nu_{\ell=0,n=27}=2363.81 \pm 3.90 \mu Hz$.
For a model slightly below $\rm 1.2 M_{\odot}$ this is consistent
with a smaller $\rm \alpha_{mlt}$ throughout the evolution.
The logarithmic derivative of the large separation
with $\rm \alpha_{mlt}$ computed from models 8 and 9
is $\rm \frac{\partial ln \Delta \nu}{\partial ln \alpha_{mlt}}=0.213$.

\section{Conclusion}\label{sec4}

In the context of the first CoRoT results, we build secular and 
dynamical models of the solar type star HD49933. We
use the CESAM stellar evolution code, the hydrodynamical code STAGGER
and the Aarhus ADIPLS oscillation package to compute the eigenfrequencies.
The purposes of this work are to calculate the impact of mass, 
composition, diffusion, 
rotation and convection on classical and seismological parameters.
We do not aim at finding the best fit to the current observational constraints
but to explore physical affects on the models.
We focus on the (average) large frequency separation
$\Delta \nu$. In particular we give the logarithmic 
derivatives of $\Delta \nu$ with mass, composition, rotation and $\rm \alpha_{mlt}$
in order to compare quantitavely their seismic influences.
Table \ref{tab7} sums up the corresponding results.
\begin{table}[Ht]
  \caption[]{Logarithmic derivatives of the large separation with various modeling parameters: mass, metal fraction, helium fraction, rotation and mixing length parameter.}
     \label{tab7}
 $$ 
     \begin{array}{lllllc}
        \hline
        \noalign{\smallskip}
        $$\rm \frac{\partial ln \Delta \nu}{\partial ln \Delta M}$$ & $$\rm \frac{\partial ln \Delta \nu}{\partial ln \Delta [O/Fe]}$$ & $$\rm \frac{\partial ln \Delta \nu}{\partial ln \Delta Y}$$ & $$\rm \frac{\partial ln \Delta \nu}{\partial ln V_{eq}}$$ & $$\rm \frac{\partial ln \Delta \nu}{\partial ln \alpha_{mlt}}$$ &  \\
        \noalign{\smallskip}
        \hline
        \noalign{\smallskip}
        2.6                & 0.83                    &    0.90                        &  $$1.1\,10^{-3}$$                                &  0.21                  &   \\
        \noalign{\smallskip}
        \hline
     \end{array}
 $$ 
\end{table}

We also investigate the derivative of the 
surface phase shift ($\beta (\nu)$) that is 
sensitive to the helium content of the star. We finally give 
the frequencies of the modes of orders n=14, n=27 and degree $\ell=0$
as they respectively minimize and maximize the surface effects on oscillations.
With one exception all the models include a core convection 
overshooting of 0.2 $ \rm H_p$. In the case of the models with rotation 
the angular momentum loss is not taken into account.
The results are as follows:
\begin{enumerate}
  \item Mass: $\Delta \nu$ is very sensitivity to mass 
(Table \ref{tab1}). It is a factor $\sim$ 3 less sensitive to composition effects and a
further factor $\sim$ 3 less sensitive to convection efficiency effects. 
The mass of HD49933 has to be tighly constrained if one wants to tackle
composition or convection efficiency issues.

  \item Composition: the assumed solar composition 
strongly influences the {\it absolute} metal content of HD49933 and therefore
its seismic properties (Table \ref{tab2}). If using seismology, a mass estimate to better than a few 
$\rm 0.01 M_{\odot}$ will not be reliable as long as the solar composition is not settled.
The derivative $\beta (\nu)$ of the surface phase shift  
depends on the helium content (Table \ref{tab2}). However the $\beta (\nu)$ dependence on mass 
shows once more that mass should be precisely constrained.

  \item  Diffusion: diffusion effects are expected
to be more important in HD49933 than in our Sun. The models taking 
diffusion into account suggest that HD49933 is more massive than the models 
that do not take diffusion into account (Table \ref{tab3}). The diffusive models rise an issue: 
when having the correct effective temperature they are too metal poor, when having the right 
metallicity they are too cool.

  \item Rotation: the rotational effects appear negligible on both classical 
and seismological parameters describing HD49933 except for the $\rm Z/X$ ratio (Table \ref{tab4}).
This point is also observed for the Sun.
The interplay between rotation and diffusion improves the situation of purely diffusive 
models by braking the gravitational settling of heavy elements.

  \item Outer convection: our hydrodynamical calculations suggest that the efficiency of convection 
in HD49933 should be slightly higher than in the Sun. This contradicts the 
better agreement of small $\rm \alpha_{mlt}$ models than large $\rm \alpha_{mlt}$ models with the observed
large frequency separation. The seismic observations of HD49933 
favor a decrease in the efficiency of convection with increasing mass.

  \item  Our best fit model has a mass of $\rm 1.17 M_{\odot}$. It includes neither 
diffusion or rotation. Its initial composition is $\rm X=0.7664$ and 
$\rm Z=5.4 \, 10^{-3}$. This mass estimate is unchanged if core
overshooting is suppressed. It is also not sensitive to the two current
possible mode identification.

\end{enumerate}

This work points out the role of sophisticated
physics in the seismological modeling of HD49933. It
stresses the need for an accurate mass and composition determination
before the observational data can be used to constrain the internal
dynamics. Moreover it shows that diffusive effects and 
their interplay with rotational effects significantly change the 
mass and age estimates of the star. Both diffusion and rotation effects
have to be taken into account in HD49933 to draw reliable conculsions.
Thus there is a need for sophisticated modelling of this CoRot target.
That is especially important as with respect to its 
temperature the star lies in the middle of the so-called lithium dip. 
Therefore the future modelling works should include the angular 
momentum loss as well as the possible
rotationally induced shear turbulence in the upper radiation
zone of the star. For the same reason it would be interesting to investigate 
the current surface abundance in lithium and beryllium from the 
observational point of view.

\begin{acknowledgements}
     We thank the anonymous referee whose remarks helped to expand and improve the content
     of this article. L. Piau is member of the UMR7158.
     This work was supported by the French
     \emph{Centre National de la Recherche Scientifique, CNRS\/} and the \emph{Centre National d'Etudes Spatiales, CNES\/}.
\end{acknowledgements}

\end{document}